# Interference Reduction in High Density WLAN Deployments using antenna Selection

Abhijeet Bhorkar and Gautam Bhanage
Aruba Networks*
1322 Crossman Ave, Sunnyvale, CA

*Abstract:* In this work, we present a novel, robust scheme for high density WLAN deployments. This scheme uses well known selection diversity at the transmitter. We show that our scheme increases the number of simultaneous transmissions at any given time without excessive overhead (compared to other schemes such as Multi-user MIMO). Furthermore, this scheme can be easily implemented using existing standards.

## 1. Introduction

Wireless LANs have become the de-facto medium for user connectivity with the internet. This typically can result in conditions with high densities (HD) of users and hence similarly more number of APs in a vicinity. Under such conditions, the performance of the network is usually limited by the interference, both between APs and across users. This proposal addresses concerns in such deployments. The following conditions make the problem worse in such HD deployments:

- **Fewer orthogonal channels:** This interference problem is aggravated by the arrival of 802.11ac where wider channel widths (up to 160MHz) will make the available number of orthogonal channels fewer. Even at 80MHz, there are only 3 orthogonal channels. Such interference effects result in reduced throughput due to contention as well as collision of packets.

- **Hidden nodes:** are fairly common scenarios [2] in WLANs where interference results in a significant number of packet collisions. In the presence of such conditions, the network performance deteriorates significantly.

- **STBC support:** Some legacy clients do not have support for STBC. In such cases, the AP is not capable of leveraging diversity gains. This can result in significant underperformance.

- **Side effect of conventional MIMO techniques:** It is well known that MIMO techniques can highly increase the reliability of the point-to-point, multi-user communication systems and increase the transmission reliability (using codes such as Space time block Codes (STBC)). While this MIMO transmission helps improve the reliability and throughput on individual links, they also result in increased interference. It would be useful to have a mechanism to counter this increased interference.
- **Alternative mechanism for MU-MIMO:** It is known that MU-MIMO can provide significant performance gain, however under perfect feedback. It is inferred that with imperfect feedback, the gains in MU-MIMO can significantly be degraded [5] undermining the gains MU-MIMO can achieve.

In this work, we propose "Max-Antenna" a software technique for MIMO to reduce the interference in the 802.11n nodes using antenna selection diversity. We show that:

*a) This approach can provide a significant improvement in performance over using plain STBC, as well reducing the performance penalty seen in scenarios with hidden nodes (cf. Section 2).*
*b) This technique can also be deployed as an alternative **robust** mechanism for MU-MIMO requiring less feedback (when the client has multiple antennas) capable of handling erroneous feedback and enabling simultaneous transmissions across AP's.*

In the sections to follow, we will provide the background for the innovation and possible implementation details.

## 2. Max-Antenna Design

Our solution using the Max-antenna framework is two pronged:
1. Using Antenna selection diversity we show how the interference in a WLAN can be reduced over the conventional approach of using STBCs.
2. Using small modifications to the MAC, we show how the max-antenna framework can be used as a substitute for MU-MIMO in HD environments. Specifically, we show how the max-antenna scheme can be used across multiple APs in an HD environment to emulate MU-MIMO. Since this approach does not actually use MU-MIMO techniques no client support is needed.

Before delving into the details of our design, we will present a brief discussion on the background for the max-antenna selection scheme.

## 2.1 Background - Antenna Selection

Antenna selection has been significantly previously studied as a mechanism to utilize the diversity and reducing the radio power by shutting of the tx/rx chains [3,4]. The objective of antenna selection has been in exploiting channel diversity and optimizing the transmission/reception power utilization on a link-to-link basis. Owing to these gains, Antenna selection capability (ASEL) has also been added in the 802.11n standard. However, until this point, antenna selection has not been considered or optimized as a mechanism to reduce to interference for a wireless networks such as wireless enterprise WLAN.

Space Time Block Codes has been a defacto method for transmission of reliable packets in current 802.11n and 802.11ac standards to provide spatial diversity gain. STBC codes provide maximum possible diversity in a MIMO system without requiring any feedback from the receiver. However, downside of this problem is that the transmitter (AP) needs to transmit over multiple antennas causing more interference to other transmissions. The objective of this work is to design a strategy which can give performance results comparable to STBC, however causing less interference to the ongoing transmissions.

Consider a transmission from AP1->Client1 with $N_t$ antennas at AP and Nr receive antennas at the client. Using STBC codes it can proved that for any transmission from AP to Client1 provides ($N_r$x$N_t$) diversity gain. In particular, the Bit error rate of the transmission is:

$$BER(AP1 -> Client1) \alpha \frac{1}{SNR^{Nr \times Nt}}$$

Note that in obtaining this BER, STBC has to transmit over multiple streams. It can be proved that if there is any ongoing transmission say from Node2->AP2, then the second transmission (say from hidden node) might be interfered significantly causing almost all the packets to drop.

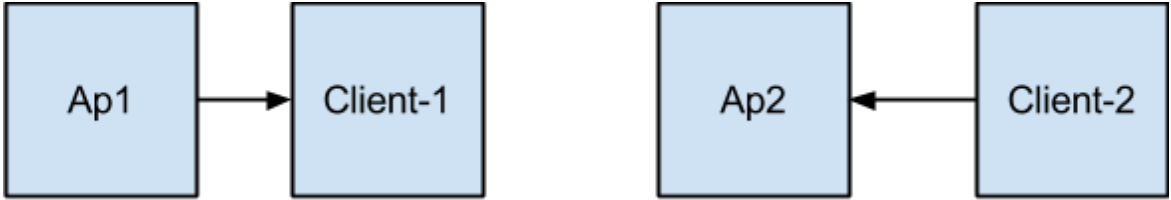

Example

The main problem with using STBC is that it can create significant interference to the surrounding transmissions.   Specifically, due the interference  caused to the Client2->AP2 transmission by AP1->Client1, Client2->Ap2 transmission cannot succeed with high probability.

Mathematically, AP2->Client2 has $N_r \times N_t$ diversity when if no other transmission is present. However, transmission from AP1 reduces the effective diversity and overall BER to

$$BER(Client2 \rightarrow Ap2) \; \alpha \; \frac{1}{SNR^{Nr \times Nt - Nt}}$$

If we use an antenna at Ap1, which has the best gain, we can minimize the interference.
Consider a channel matrix $H = [h_1 \; h_2 .. \; h_{nt}]$ from AP1, when Nt is number of antennas at AP1.
 We choose the antenna at AP1 with maximum channel gain.

$$Antenna \; = \; argmax_i \; (||hi||), \quad \text{where } ||h_i|| \text{ is the 2-norm.} \quad (1)$$

Since, number of antennas used for the transmission at AP1 are less (actual only 1) the BER at AP2 will be significantly improved.

$$BER(Client2 \rightarrow Ap2) \; \alpha \; \frac{1}{SNR^{Nr \times Nt - 1}}$$

## 2.2 Max-Antenna ASEL

Implementing **MaxAntenna** needs information about the best antenna use.  We can utilize the feedback mechanisms defined by Antenna Selection Capability (ASEL).  This feedback contains sufficient information to obtain the best antenna using implicit and explicit feedback mechanisms. However, typically, ASEL is not supported on access points/clients. Hence, along with these the feedback approach, we also propose that using channel reciprocity we can obtain the best antenna with no feedback.

**Feedback based implementation:**
The 802.11n defines Antenna Selection Capability as a optional feature to select best antenna in the device when the number of antennas are larger than number of radios. ASEL thus provides a form of selection diversity by intelligently selecting the antennas.
ASEL sounding packets are used to pick Rx and Tx antennas between two devices.
which has best channel conditions and yield highest signal to interference ratio. Note that to implement ASEL based solution, both AP and STA should have Antenna Selection Capability enabled. ASEL allows to obtain the best antennas for RX/TX after sounding is complete (using CSI action frame or Indices feedback action frame).

**No Feedback implementation:**
Interestingly, using the channel reciprocity, we propose a simple mechanism without any feedback overhead.

Precisely, for each transmission at the AP, we can obtain the received signal strength for its acknowledgement. The RSSI value is stored for each antenna in the *ts* descriptor populated in the driver. The RSSI value is a proxy for the term $\|h_i\|$ in Equation 1.
Thus, antenna with largest RSSI is selected to be the best antenna.

In atheros driver, ts->ts_rssi_ctl0, ts->ts_rssi_ctl1, ts->ts_rssi_ctl2 denote the RSSI values in the transmit descriptor. Thus, in the driver, we can select the antenna with largest ts->ts_rssi_ctl.

## 2.3 Simultaneous Transmissions in HD:

The **MaxAntenna** scheme can be used to improve the throughput. It is based on following fact from [1]
**Fact 1:**
If the clients have number of antennas equalling to that of AP's (AP and clients have N antennas), it is possible to transmit from N AP's simultaneously with high reliability.

 802.11 is a CSMA/CA based protocol and does not allow simultaneous transmissions from multiple AP's. To allow for simultaneous transmissions, CSMA based MAC need to be changed. Possible generic implementation is suggested in [1].

Next, we propose an implementation of modified MAC for controller based enterprise WLAN.
For the easy of deposition, we assume N=4 antennas at each AP and client in the wireless network.

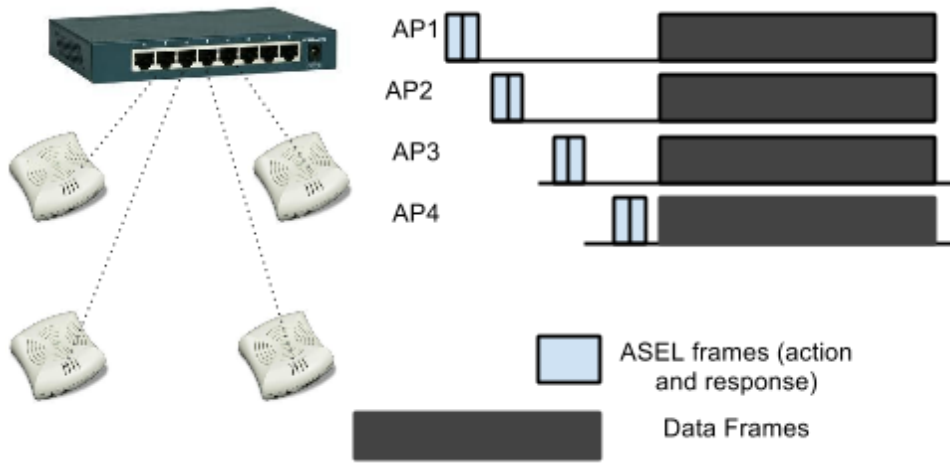

The aim of the scheme is to enable and empower the wlan system with simultaneous transmissions from different APs.

1. In the first step, the controller first gathers information if there is downlink transmission for a set of APs denoted by $G$, $|G|=N$ and indexed priority by the controller.
2. AP $\in G$ indexed as one, initiates ASEL transmission. Based on the ASEL action frame, its corresponding client feedback with the corresponding best antenna. After This process is iterated for all APs after previous AP finishes its action-response sequence.
3. Finally, all APs transmit the corresponding data packets.

**Overhead :**
   Max-antenna method employs a feedback mechanism and incurs some overhead.
However, we argue that the overhead is small compared to the gains obtained.
Assuming that the total transmission time for action-response sequence is 1ms [5] and assuming that TxOP is set to 4ms (typical), we can achieve the transmission in around 5ms (compared to around 16ms without using Max-Antenna scheme which needs sequential transmissions from AP).
   Overhead for the ASEL frames can further be reduced if some apriori knowledge is available from previous history.

## 3. Preliminary Evaluation

In this Section, we evaluate the performance of Antenna selection diversity vs. STBC codes. We show that under interfering environment, selection diversity outperforms alamouti codes.

Simulation Setup:
  We have conducted simulations in matlab to model the effect of interference on a particular link (AP-Client) due to ongoing transmissions . We assume that the AP and clients have 4 antennas. We have assumed that the channel between each AP and client is rayleigh fading with unit norm. We have plotted the performance of Antenna selection diversity and STBC codes in presence of 3 ongoing transmission hidden from transmission on focus when BPSK modulation is used. Figure 1 plots the Bit error probability of STBC codes and Selection diversity as SNR is varied by changing the power at the AP. In Figure 2, we repeat the experiments assuming only 3 simultaneous transmissions. Note that, in the second case, due to less interference, the performance of STBC and MaxAntenna have improved and effective gain between these two techniques have reduced.

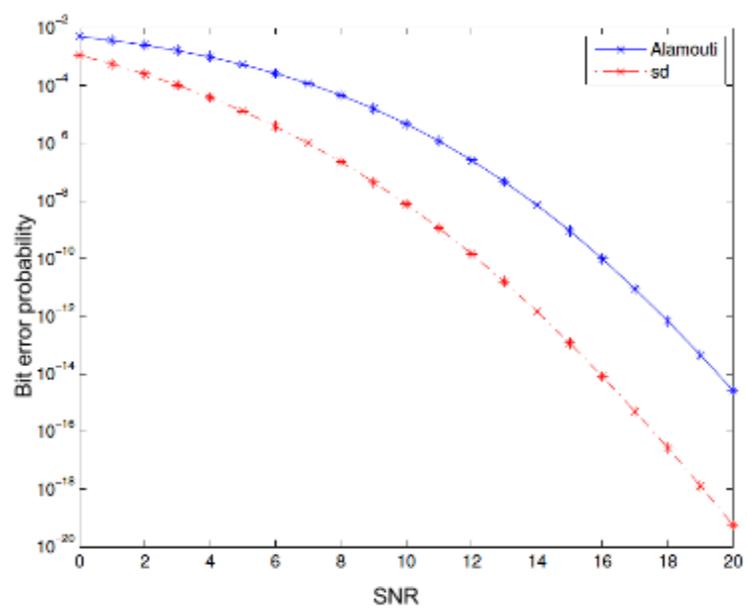

Figure 1

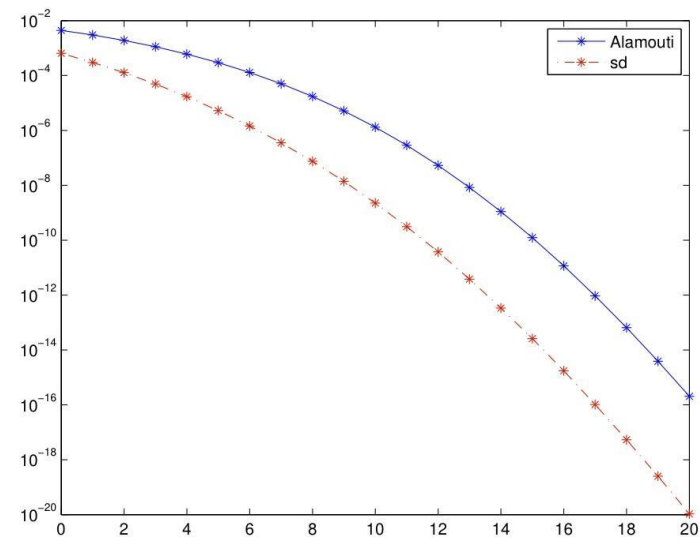

Figure 2

Note that in high density deployments, typically SNR is high. Figure 1 shows that for MaxAntenna effective Bit error is significantly lower than STBC. For typical SNR values, we observe that BER is $10^{-6}$ and $10^{-4}$ for selection diversity and STBC respectively. This leads to Packet error rate of 0.4% and 33% respectively. This large difference between the PER suggests the ability of superiority of selection diversity in wlan networks and its superiority in presence of hidden nodes. Furthermore, in 802.11ac based WLAN networks, Figure 1 also shows that simultaneous transmission are feasible in a network with clients supporting multiple antennas.

## 4. Conclusion

We have proposed a simple method to tackle interference in the network by selecting the *best* antenna during transmission. It has high potential to reduce the collisions for hidden node terminals. Further, we also proposed mechanism for 802.11ac by allowing simultaneous transmission from multiple APs on the same channel.